\documentclass[superscriptaddress, twocolumn, prl, longbibliography]{revtex4-1}
\usepackage{amsmath, amssymb, color, graphicx, tikz}
\usepackage[utf8]{inputenc}
\definecolor{linkcolor}{rgb}{0,0,0.6} 
\usepackage[pdftex,colorlinks=true,
	pdfstartview = FitV,
	linkcolor    = linkcolor,
	citecolor    = linkcolor,
	urlcolor     = linkcolor,	
	hyperindex   = true,
	hyperfigures = false]{hyperref}

\newcommand{\resub}[1]{{#1}}

\begin{document}

\title{Optimal power and efficiency of odd engines}

\author{\'Etienne Fodor}
\affiliation{Department of Physics and Materials Science, University of Luxembourg, L-1511 Luxembourg}

\author{Anton Souslov}
\affiliation{Department of Physics, University of Bath, Claverton Down, Bath BA2 7AY, UK}

\begin{abstract}
	Odd materials feature antisymmetric response to perturbations. This anomalous property can stem from the nonequilibrium activity of their components, which is sustained by an external energy supply. These materials open the door to designing innovative engines which extract work by applying cyclic deformations, without any equivalent in equilibrium. Here, we reveal that the efficiency of such energy conversion, from local activity to macroscopic work, can be arbitrarily close to unity when the cycles of deformation are properly designed. We illustrate these principles in some canonical viscoelastic materials, which leads us to identify strategies for optimizing power and efficiency according to material properties, and to delineate guidelines for the design of more complex odd engines.
\end{abstract}

\maketitle


The aim of engine design is not only practical, but also conceptual: Studying thermal engines was pivotal for the development of equilibrium thermodynamics~\cite{Callen}. The design of minimal cycles, whose performances can be computed exactly, has led to recipes that optimize more complex engines in terms of universal observables (\textit{i.e.}, power and efficiency). Indeed, the seminal Carnot cycle, which places a simple bound on the efficiency of \textit{any} thermal engine~\cite{Carnot}, still serves today as a testbed to guide challenges in modern research, such as the design of micrometer-scale engines~\cite{Blickle2011, Sood2016, Martinez2016, Martinez2017}.

Materials which evade equilibrium constraints offer opportunities to devise innovative engines with unprecedented performances. Active matter encompasses nonequilibrium systems where every unit has internal machinery powering its motion~\cite{Marchetti2013, Bechinger2016, Marchetti2018}. Active systems are either living (\textit{e.g.}, bacterial swarms~\cite{Libchaber2000, Elgeti2015}), social (\textit{e.g.}, animal groups~\cite{Cavagna2014}), or synthetic (\textit{e.g.}, catalytic colloids in fuel bath~\cite{Bechinger2013, Palacci2013}). Each unit can be regarded as a \textit{microscopic engine} converting the energy fuel stored in the environment (\textit{e.g.}, nutrients feeding bacteria) into autonomous motion. A natural question, which has already received extensive attention~\cite{Zakine2017, Martin2018, Pietzonka2019, Saha2019, Szamel2020, Park2020, Holubec2020, Ekeh2020, Puglisi2021, Stark2020, Cates2021}, is how to exploit individual self-propulsion to design \textit{macroscopic active engines}: How to harvest particle-based, disordered motion to produce macroscopic, sustained motion?

In most active systems, the metabolic rate characterizing microscopic fuel conversion is constant, and active units are \textit{always} self-propelling. In such liquid materials, the energy stemming from fuel conversion sustains self-propulsion and becomes dissipated in the surrounding thermostat, with negligible contribution to macroscopic work extraction. As a consequence, active fluids naturally yield engines with very low efficiency~\cite{Ekeh2020}. However, in principle, units need not always be actively moving, instead their activity may be modulated in proportion to external perturbations.

In search for materials with such adaptive units, the motivation is to design protocols by \textit{minimizing} the individual dissipation and \textit{maximizing} the macroscopic work. Importantly, this approach focuses on optimizing the macroscopic energy conversion irrespective of the details of microscopic fuel conversion. We ask: How does one properly interface with active constituents given minimal assumptions on their individual dynamics?

\begin{figure}[b]
	\centering
	\includegraphics{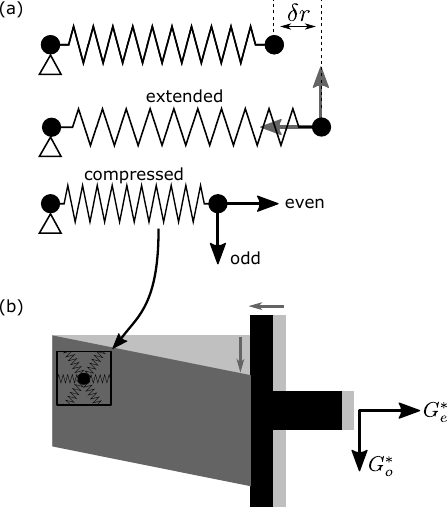}
	\caption{\resub{Schematic of odd engines at microscopic and macroscopic scales.
	(a) A ball-and-spring microscopic model. The restoring and dissipative forces on the spring are captured by the typical (even) part of the complex modulus $G^{*}_{e}$. Analogously, the transverse (odd) forces are captured by the odd complex modulus $G^{*}_{o}$. The sign of the odd and even forces depends on whether the spring is extended (middle) or compressed (bottom).
	(b) Continuum descriptions of odd engines. For a macroscopic material composed of springs in part (a), $G^*_{e}$ describes the viscoelastic response of passive materials, whereas the odd response $G^*_{o}$ couples, for example, a shear stress to a different shear strain.}
		}
	\label{fig1}
\end{figure}

Odd materials are novel nonequilibrium systems with antisymmetric relations between stress and strain, \resub{see Fig.~\ref{fig1}}. Their components can either be subject to constant torques (\textit{e.g.}, colloids rotating with an external magnetic field~\cite{Aranson2017, Bartolo2019}), or trigger active internal forces under external perturbations (\textit{e.g.}, metabeam made of piezoelectric patches with electronic feedback~\cite{Chen2020}). These materials then acquire some anomalous mechanical properties: \textit{Odd viscosity} (equivalently, Hall viscosity~\cite{avron1995viscosity, Banerjee2017}) has been reported in models of spinning particles~\cite{Nguyen2014, Bartolo2016}; \textit{Odd elasticity} has been studied in models of elastic networks, whose bonds yield local transverse forces upon compression and/or extension~\cite{Scheibner2020a, Coulais2021}. In both cases, the response non-reciprocally couples different changes in shape, which can lead, for instance, to compression as a result of applied torque~\cite{Banerjee2017, Scheibner2020a}.

For our purpose, odd elasticity stands out as the key ingredient in designing efficient engines. Indeed, this response typically arises in assemblies of transducers, acting as both sensors and actuators, which adapt their activity to external cues~\cite{Scheibner2020a, Chen2020, Coulais2021}. Such materials have been used to design nonequilibrium engines with slow cyclic deformations~\cite{Scheibner2020a}. The engines exploit the fact that the work now depends on the \textit{whole} path of deformations (\textit{i.e.}, not only on initial and final points), in contrast to thermal engines~\cite{Callen}. A remaining challenge is to explore whether these engines indeed fulfill the promise of efficient and powerful energy conversion from local activity, leading to macroscopic work extraction.

In this paper, we examine how to optimize the performances of odd engines using a continuum theory of odd materials. We address solids that are well modeled by a linear viscoelastic response: The elasticity and viscosity tensors both depend on the frequencies of applied perturbation. In general, we show that (i)~if odd elasticity is strong at low frequencies, then slow cycles are always advantageous, in which case there is a trade-off between power and efficiency analogously to thermal engines, whereas (ii)~if the elasticity dies out at low frequencies, then fast cycles are advantageous. These cases are exemplified by canonical models imported from materials science, such as the Kelvin-Voigt solid, the standard linear solid, and the power-law solid. For each model, we explore how the cycle needs to be tailored to material specificities to optimize power and/or efficiency. Overall, our results provide intuition for how to best exploit odd properties, and illustrate some concrete guidelines for the design of more complex engines.


We consider a linear viscoelastic material~\cite{Rogers1963, Day1971, Fabrizio1985}, whose constitutive relation between stress $\sigma$ and strain $u$ (or, more generally, displacement gradient) is given in the Fourier domain in terms of the dynamic modulus $G^*$:
\begin{equation}\label{eq:cons}
	\tilde\sigma_{ij}(\omega) = G^*_{ijkl}(\omega) \, \tilde u_{kl}(\omega) ,
\end{equation}
where $\omega$ is the frequency, and we use implicit summation over repeated indices throughout. The storage ($G'$) and loss ($G''$) moduli correspond, respectively, to the real and imaginary parts of $G^*$. The (novel, antisymmetric) odd parts $G_{o}$ and the (standard, symmetric) even parts $G_{e}$ of each modulus obey $G_{o,ijkl}=-G_{o,klij}$ and $G_{e,ijkl}=G_{e,klij}$, \resub{see Fig.~\ref{fig1}}. For a protocol varying the strain $u$ periodically, the work $\cal W$ produced during a cycle of period $\tau$ reads
\begin{equation}\label{eq:work}
	{\cal W} = - \int_0^\tau \sigma_{ij}(t) \dot u_{ij}(t) dt .
\end{equation}
Within our convention, work is extracted from the material when ${\cal W}>0$. Combining~(\ref{eq:cons}-\ref{eq:work}), the work can be written in terms of the strain Fourier coefficients $\hat u_n$ as
\begin{equation}\label{eq:work_g}
	{\cal W} = 4\pi \sum_{n>0} n \, \hat u_{-n,ij} \, \hat u_{n,kl} \, \big[ \text{i} G'_{o,ijlk} (\omega_n) - G''_{e,ijlk} (\omega_n) \big] ,
\end{equation}
where $\omega_n = 2\pi n/\tau$, and we have used the fact that $G'$ is even (and $G''$ is odd) with respect to $\omega$. The power is the work per cycle period, ${\cal P}={\cal W}/\tau$, and the quasistatic work ${\cal W}_{qs}$ follows from Eq.~\eqref{eq:work_g} by taking the limit of large $\tau$ (\textit{i.e.}, $\omega_n\to0$). The dissipated energy $\cal D$ is defined as the part of the work associated with the loss modulus:
\begin{equation}
	{\cal D} = 4\pi \! \sum_{n>0} \! n \, \hat u_{-n,ij}  \, \hat u_{n,kl} \, G''_{e,ijlk} (\omega_n) ,
	\quad
\end{equation}
and, inspired by previous works on monothermal protocols~\cite{Prost1999, Cates2021}, we use the following measure of efficiency:
\begin{equation}\label{eq:heat}
	{\cal E} = \frac{\cal W}{{\cal W} + {\cal D}} .
\end{equation}
Interestingly, energy-conserving features (\textit{i.e.}, odd viscosity and even elasticity) affect neither efficiency nor power. Since $\cal D$ must be positive to ensure material stability, $\cal E$ lies in the range $(-\infty,1]$. For strictly dissipative materials (${\cal W}+{\cal D}=0$), any cycle has $\cal E = -\infty$, whereas for other materials, cycles can support $\cal E$ close to $1$ whenever dissipation becomes negligible (${\cal W}\gg{\cal D}$). Note that our definition of efficiency addresses how the energy input due to odd features can be transduced into extracted work, without taking into account the underlying microscopic mechanisms which sustain this energy input. When dissipation arises from collisions with molecules of the surrounding thermostat, $\cal D$ is the heat absorbed by the thermostat~\cite{Lebon1986, Morton1991}. In other contexts, such as in granular materials with non-elastic collisions between system components~\cite{Han2020}, or in epithelial tissues with dry friction due to the substrate~\cite{Bi2015, Bi2016}, $\cal D$ can be distinct from heat.

With the definitions in Eqs.~(\ref{eq:work_g}--\ref{eq:heat}), the sum of work and dissipated energy, ${\cal W}+{\cal D}$, does not reduce to the boundary term $[\sigma_{ij}u_{ij}]_0^\tau$, as would be the case in the absence of any energy source~\cite{Morton1991}. Instead, the sum ${\cal W}+{\cal D}$ is generally non-zero, even if the protocol is periodic, due to the odd storage modulus $G'_o$. This illustrates that some energy must be supplied externally to the material~\cite{Lebon1986} to sustain its odd properties~\cite{Scheibner2020a}. This situation is reminiscent of engines composed of self-propelled particles, where the energy balance ${\cal W}+{\cal D}$ includes explicitly the cost of microscopic self-propulsion~\cite{Cates2021}. Yet, in contrast with self-propelled particles, some odd materials behave as static solids, which do not dissipate any energy when they are at rest. As we shall see, this distinction can lead to nominal increases in $\cal E$, when the cycle is appropriately designed, with respect to previous active engines~\cite{Ekeh2020, Cates2021}.

\begin{figure*}T
	\centering
	\includegraphics[width=\linewidth]{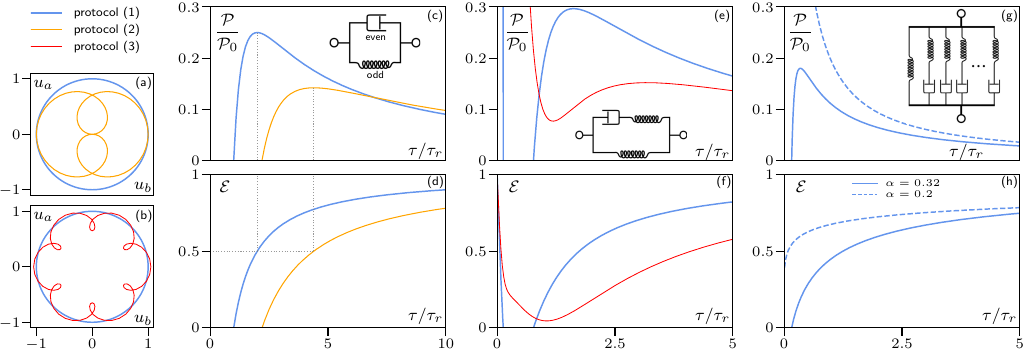}
	\caption{Performances of odd materials under various protocols.
		(a-b)~Three cyclic protocols in strain space $\{u_a,u_b\}$ with same quasistatic work (${\cal W}_{qs}=2\pi K\sum_n \! n a_n^2$): (1)~circular protocol, $a_n=\delta_{n,1}$, (2)~non-circular protocol, $a_n=(1/2)(\delta_{n,1}+\delta_{n,3})$, and (3)~non-circular protocol, $a_n=(1/5)(3\sqrt{2}\delta_{n,1}+\delta_{n,7})$, see Eq.~\eqref{eq:prot}.
		(c-h)~Power and efficiency, respectively $\cal P$ and $\cal E$, as functions of the cycle time $\tau$, where $\tau_r$ is a relaxation time scale and ${\cal P}_0={\cal W}_{qs}/\tau_r$ is a reference power. We compare these metrics $\{{\cal P},{\cal E}\}$ for three materials subject to the protocols in (a-b). The material mechanics is characterized by the odd storage modulus $g'$ and even loss modulus $g''$ (see Eq.~\eqref{eq:mod}): (c-d)~odd Kelvin-Voigt (KV) solid (Eq.~\eqref{eq:kv_mech}), (e-f)~odd standard linear (SL) solid (Eq.~\eqref{eq:sl}), and (g-h)~odd power-law (PL) solid (Eq.~\eqref{eq:pl}). The insets display some schematic representations of each material in terms of even/odd dampers and springs.
		(c-d)~KV solids exhibit a trade-off between power (maximum at intermediate $\tau$) and efficiency (maximum at large $\tau$). The efficiency at maximum power is always $1/2$ independently of the protocol details, see dotted lines. Non-circular protocols systematically reduce both the efficiency at all times and the maximum power.
		(e-f)~SL solids have diverging power at small $\tau$, and maximum efficiency at both small and large $\tau$. The engine cannot extract work in an intermediate regime of $\tau$ (where ${\cal P}<0$ and ${\cal E}<0$) for circular protocols (blue line), yet it can extract work at all $\tau$ for some non-circular protocols (red line).
		(g-h)~PL solids have either diverging power at small $\tau$ (when $\alpha<1/4$), or maximum power at intermediate $\tau$ (when $\alpha>1/4$). The efficiency is always maximum at large $\tau$.
		Parameters: $K=1$, $\tau_r=1$, $\lambda=0.1$.
		}
	\label{fig2}
\end{figure*}

In what follows, for simplicity, we consider cases where $G^*$ reduces to a matrix in the two-dimensional strain subspace $\{u_a,u_b\}$:
\begin{equation}\label{eq:mod}
	G^* =
	\begin{bmatrix}
		\text{i}g'' & g'
		\\
		- g' & \text{i}g''
	\end{bmatrix} + G'_e + \text{i} G''_o
	,
\end{equation}
where $g'$ embodies the odd components of the storage modulus, and $g''$ the even components of the loss modulus. Although minimal, this choice captures the essential ingredients at play in odd materials. Decomposing the strain protocol as (see Figs.~\ref{fig2}(a-b))
\begin{equation}\label{eq:prot}
	\big[u_a, u_b\big] = \sum_{n>0} a_n \,\big[\cos(\omega_n t), \sin(\omega_n t)\big] ,
\end{equation}
we deduce power and efficiency from Eqs.~(\ref{eq:work_g}--\ref{eq:heat}) and~\eqref{eq:prot}:
\begin{equation}\label{eq:gen}
	{\cal P} = \sum_{n>0} \omega_n a_n^2 (g'- g'')(\omega_n) ,
	\quad
	{\cal E} = \frac{\cal P}{\sum_n \omega_n a_n^2 g'(\omega_n)} .
\end{equation}
Work can be extracted (${\cal P}>0$) only if the material behaves as an odd solid ($g'>g''$, \textit{i.e.}, storage greater than loss) in a finite range of frequencies. This criterion allows one to rule out some systems, such as the chiral fluid described in Ref.~\cite{Han2020}, as unfit for work extraction. For cycles with a single harmonic ($a_n\propto\delta_{n,\alpha}$), the corresponding frequency $\omega_\alpha$ must satisfy $g'(\omega_\alpha)>g''(\omega_\alpha)$, which defines the appropriate frequency range for work extraction. The efficiency reduces to $1-g''(\omega_\alpha)/g'(\omega_\alpha)$ and can get arbitrary close to unity by increasing the ratio of odd storage to even loss moduli. In general, combining several harmonics can potentially help ensure work extraction and increase efficiency at any cycle time $\tau$, by selecting the values $a_n$ according to material parameters.


Equipped with the expressions~\eqref{eq:gen}, we now proceed by analyzing some familiar types of rheology. First, we address the case of odd Kelvin-Voigt (KV) solids:
\begin{equation}\label{eq:kv_mech}
	g'(\omega)=K, \qquad g''(\omega)=\text{i}\omega \eta ,
\end{equation}
where $K$ is the odd elastic modulus, and $\eta$ the viscosity (see inset of Fig.~\ref{fig2}(c)). These materials behave as even viscous fluids at short times and odd elastic solids at long times. Such a material could be constructed from an elastic network of active bonds embedded in a viscous fluid, with cycles in the space of shear strains~\cite{Scheibner2020a, Banerjee2021}. The associated power and efficiency have simple expressions:
\begin{equation}\label{eq:kv}
	{\cal E} = \frac{\tau\cal P}{{\cal W}_{qs}} = 1 - \frac{\bar\tau}{\tau} ,
	\qquad
	{\cal W}_{qs} = 2 K \iint du_a du_b .
\end{equation}
The engine extracts work at all times larger than $\bar\tau$, where $\bar\tau$ depends on $a_n$ and $\tau_r=\eta/K$. The power has a non-monotonic behavior with peak value at $2\bar\tau$ and vanishes at long times, see Fig.~\ref{fig2}(c), analogously to thermal engines~\cite{Seifert2008, Esposito2010} and monothermal cyclic engines with self-propelled particles~\cite{Ekeh2020}. Interestingly, the maximum power $\bar{\cal P}$ is proportional to the ratio of the squared area in strain space $\{u_a,u_b\}$, namely $(\iint du_a du_b)^2$, over $\int_0^1 (\dot u_a^2+\dot u_b^2) ds$, where $\dot u_a=du_a/ds$. For a given ${\cal W}_{qs}$, optimizing $\bar{\cal P}$ then requires minimizing the perimeter at fixed area: This is achieved for a circular protocol. Adding any higher harmonics systematically reduces both efficiency $\cal E$ for all times and maximum power $\bar{\cal P}$, see Figs.~\ref{fig2}(c-d). If $u_a$ and $u_b$ are associated with different viscosities, respectively $\eta_a$ and $\eta_b$ (\textit{e.g.}, for rotation and dilation~\cite{Scheibner2020a}), the optimal protocol now describes an ellipse: $[u_a,u_b] \propto [c \cos(\omega_1 t), c^{-1} \sin(\omega_1 t)]$ where $c = (\eta_b/\eta_a)^{1/4}$. In addition, all these results still hold for non-linear elasticity, namely when ${\cal W}_{qs} = 2 \iint K(u_a,u_b) du_a du_b$, as long as the even loss modulus remains purely viscous.

The efficiency of KV solids increases monotonically with the cycle time and converges to unity (see Fig.~\ref{fig2}(d)), in contrast with both cyclic active engines, where $\cal E$ vanishes at long times~\cite{Ekeh2020}, and thermal engines, where $\cal E$ is bounded by the Carnot efficiency~\cite{Carnot}. This illustrates how odd engines typically outperform other engines, either thermal or active, when they behave purely as odd elastic solids (at long times for KV solids). As for thermal engines~\cite{Seifert2008, Esposito2010}, there is a trade-off in KV solids between achieving either maximal power (at intermediate times) or high efficiency (at long times). Surprisingly, the efficiency at maximum power is universal for KV solids and always equals $1/2$ independently of both material properties ($K,\eta$) and cycle details ($a_n$).


In search of odd materials achieving simultaneously maximal power and high efficiency, we now address odd standard linear (SL) solids. For simplicity, we consider that $g'$ and $g''$ are related by (see inset of Fig.~\ref{fig2}(e))
\begin{equation}\label{eq:sl}
	g'(\omega)+\text{i}g''(\omega) = K \frac{1+\text{i}\omega\tau_r}{1+\text{i}\lambda\omega\tau_r} ,
\end{equation}
where $\tau_r$ is a relaxation time, and the dimensionless parameter $\lambda$ obeys $0<\lambda<1$. The odd SL solids behave as odd elastic solids at both short and long times. In general, considering a schematic representation of a material with dampers and springs, each one being either odd or even, $g'$ and $g''$ can be expressed in terms of separate viscosities and elastic moduli~\cite{Banerjee2021}. SL solids with even moduli correspond, for instance, to the mechanics of vertex models~\cite{Tong2021}, which describe epithelial tissues as a dense assembly of self-propelled particles~\cite{Bi2015, Bi2016}. Odd components might arise in such a model when considering self-propelled particles with chirality~\cite{Liebchen2017}.

The quasistatic work is analogous to that of odd KV solids~\eqref{eq:kv}; $\cal P$ and $\cal E$ directly follow from Eqs.~\eqref{eq:gen} and~\eqref{eq:sl}, see footnote~\footnote{Performances of odd standard linear solids: power ${\cal P} = K \sum_n \omega_n a_n^2 \frac{1 + (\lambda-1)\omega_n\tau_{\rm r} + \lambda(\omega_n\tau_{\rm r})^2}{1 + (\lambda\omega_n\tau_{\rm r})^2}$, and efficiency ${\cal E} = {\cal P} \; \Big[ K \sum_n \omega_n a_n^2 \frac{1+\lambda(\omega_n\tau_{\rm r})^2}{1+(\lambda\omega_n\tau_{\rm r})^2} \Big]^{-1}$.}. For $\lambda>3-2\sqrt{2}$, we get from Eq.~\eqref{eq:sl} that $g'>g''$ at all frequencies, so that the power is always positive (see Eq.~\eqref{eq:gen}): The cycle extracts work for any cycle time $\tau$. In contrast, for $\lambda<3-2\sqrt{2}$, considering protocols with a single harmonic leads to an intermediate range of $\tau$ without work extraction. Yet, combining several harmonics can still yield work extraction at all times, see Figs.~\ref{fig2}(e-f). For any $\lambda$, the power vanishes at long times, and it diverges as $\tau^{-1}$ at short times toward large positive values. The efficiency converges to unity at short and long times, since odd SL solids behave as odd elastic solids in these regimes. Therefore, in contrast with odd KV solids, the engine can now produce high $\cal P$ with $\cal E$ arbitrary close to unity, when operated at very short times: There is no longer any trade-off between power and efficiency. Interestingly, a similar behavior is observed for materials which act as odd solids at short times, without necessarily being solids at long times, such as odd Maxwell liquids~\cite{Banerjee2021}.


As a final illustration, we consider odd power-law (PL) solids, which combine features of KV and SL solids:
\begin{equation}\label{eq:pl}
	g'(\omega)+\text{i}g''(\omega) = K \big[1+(\text{i}\omega\tau_r)^\alpha\big] ,
\end{equation}
where $0<\alpha<1/2$. Such materials behave as odd elastic solids at long times, and reduce to odd KV solids when $\alpha=1/2$. When $\alpha<1/4$, although these materials are not purely odd elastic solids at short times, the storage modulus is still larger than the loss. PL mechanics (without any odd component) are observed when probing the intracellular environment of some living systems with microrheology~\cite{Hoffman2006, Gallet2006, Ahmed2016, Ahmed2018}. In general, power-law behavior emerges when considering a large assembly of parallel dampers and springs~\cite{Basar}, see inset of Fig.~\ref{fig2}(g), so that the power-law relaxation is approximated by a series of exponential relaxations, known as the Prony series~\cite{Soussou1970}.

The quasistatic work is the same as for odd KV and SL solids, see Eq.~\eqref{eq:kv}. We straightforwardly deduce $\cal P$ and $\cal E$ from Eqs.~\eqref{eq:gen} and~\eqref{eq:pl}, see  footnote~\footnote{Performances of odd power-law solids: power ${\cal P} = K \sum_n \omega_n a_n^2 \big[ 1 + (\cos(\alpha\pi) - \sin(\alpha\pi)) (\omega_n\tau_{\rm r})^{2\alpha} \big]$, and efficiency ${\cal E} = {\cal P} \; \Big\{ K \sum_n \omega_n a_n^2 \big[1 + \cos(\alpha\pi)(\omega_n\tau_{\rm r})^{2\alpha} \big] \Big\}^{-1}$.}. The exponent $\alpha$ controls the transition between when the PL solid behaves like an odd KV solid (at large $\alpha$) and when it behaves like an odd SL solid (at small $\alpha$), see Figs.~\ref{fig2}(g-h). For $\alpha<1/4$, the mechanics in Eq.~\eqref{eq:pl} is such that $g'>g''$ for all $\omega$, yielding work extraction at any cycle time $\tau$, and the power diverges at short times as for odd SL solids. For $\alpha>1/4$, work is extracted at times larger than a threshold value, which depends on the parameters $\tau_r$, $a_n$, and $\alpha$, and the power has a non-monotonic behavior analogous to odd KV solids. In both cases, $\cal E$ is monotonic and converges to unity at large times, as for odd KV solids, yet the efficiency at maximum power for $\alpha>1/4$ is no longer universal, in contrast with odd KV solids. Although engines with mechanics in Eq.~\eqref{eq:pl} cannot reach simultaneously both high power $\cal P$ and efficiency $\cal E$ close to unity, this combination may be achieved by considering independent PL behaviors for $g'$ and $g''$.


In this paper, we put forward a systematic framework to predict and optimize the power and efficiency of odd engines. We consider a series of canonical odd materials and compare their performances to illustrate some generic features. Thus, we reveal that the efficiency gets arbitrarily close to unity when materials behave as odd elastic solids. The crucial difference compared to cyclic engines made of dilute self-propelled particles, whose efficiency is very low~\cite{Ekeh2020}, is that solids do not dissipate energy at rest. Consequently, operating the cycle slowly is always a good strategy for efficient work extraction when the materials behave as odd \textit{static} solids. Since the power vanishes at long times, there is typically a trade-off between efficiency and power, which is reminiscent of thermal engines~\cite{Seifert2008, Esposito2010}, although it can be circumvented if materials behave as odd solids at short times.

Importantly, our efficiency does not account for any dissipation due to the underlying active components that give rise to odd elasticity, which would reduce the overall efficiency. For instance, the transverse forces of active bonds in elastic networks can stem from internal propellers, activated under compression/extension~\cite{Scheibner2020a}. Interestingly, optimizing the conversion of some energy resource (\textit{e.g.}, local battery) into odd elasticity is an issue separate from the engine optimization addressed here. In that respect, our results already provide some insightful perspectives on how to best convert odd elasticity into useful work. Our framework is directly relevant to guide the design of future odd materials, with a view to extracting work at maximum power and efficiency.


\acknowledgements{
We acknowledge insightful discussions with Colin Scheibner and Vincenzo Vitelli, and also other illuminating discussions throughout the virtual 2020 KITP program on ``Symmetry, Thermodynamics and Topology in Active Matter'', which was supported in part by the National Science Foundation under Grant No. NSF PHY-1748958. \'E.F.~acknowledges support from an ATTRACT Investigator Grant of the Luxembourg National Research Fund. A.S.~acknowledges the support of the Engineering and Physical Sciences Research Council (EPSRC) through New Investigator Award No.~EP/T000961/1 and of the Royal Society under grant No.~RGS/R2/202135.
}


\bibliography{References.bib}

\end{document}